\documentclass[compsoc,conference,a4paper,10pt,times]{IEEEtran}
\IEEEoverridecommandlockouts
\usepackage{cite}
\usepackage{amsmath,amssymb,amsfonts}
\usepackage{algorithmic}
\usepackage{graphicx}
\usepackage{textcomp}
\usepackage{bmpsize}
\usepackage{xcolor}
\usepackage{lipsum}
\usepackage[colorlinks=true,urlcolor=black]{hyperref}
\usepackage{xurl}
\usepackage{booktabs} 
\usepackage{tabularx} 
\usepackage{array}
\usepackage{parskip}
\usepackage{multirow}

\def\BibTeX{{\rm B\kern-.05em{\sc i\kern-.025em b}\kern-.08em
    T\kern-.1667em\lower.7ex\hbox{E}\kern-.125emX}}

\begin{document}

\title{How Effective Are Publicly Accessible Deepfake Detection Tools? A Comparative Evaluation of Open-Source and Free-to-Use Platforms}

\author{
    \IEEEauthorblockN{Michael Rettinger\textsuperscript{1}\IEEEauthorrefmark{1}, 
    Ben Beaumont\textsuperscript{1}\IEEEauthorrefmark{1}, Nhien-An Le-Khac\IEEEauthorrefmark{2}, and 
    Hong-Hanh Nguyen-Le\IEEEauthorrefmark{1}}
    \IEEEauthorblockA{\IEEEauthorrefmark{1}School of Computer Science,
    University College Dublin, Dublin, Ireland\\
    \{michael.rettinger, ben.beaumont, hong-hanh.nguyen-le\}@ucdconnect.ie}
    \IEEEauthorblockA{\IEEEauthorrefmark{2}School of Computer Science,
    University College Dublin, Dublin, Ireland\\
    an.lekhac@ucd.ie}
    \thanks{\textsuperscript{1}These authors contributed equally to this work.}
}

\maketitle

\begin{abstract}
    The proliferation of deepfake imagery poses escalating challenges for practitioners tasked with verifying digital media authenticity. While detection algorithm research is abundant, empirical evaluations of publicly accessible tools that practitioners actually use remain scarce. This paper presents the first cross-paradigm evaluation of six tools, spanning two complementary detection approaches: forensic analysis tools (InVID \& WeVerify, FotoForensics, Forensically) and AI-based classifiers (DecopyAI, FaceOnLive, Bitmind). Both tool categories were evaluated by professional investigators with law enforcement experience using blinded protocols across datasets comprising authentic, tampered, and AI-generated images sourced from DF40, CelebDF, and CASIA-v2. We report three principal findings: forensic tools exhibit high recall but poor specificity, while AI classifiers demonstrate the inverse pattern; human evaluators substantially outperform all automated tools; and human-AI disagreement is asymmetric, with human judgment prevailing in the vast majority of discordant cases. We discuss implications for practitioner workflows and identify critical gaps in current detection capabilities.
\end{abstract}

\begin{IEEEkeywords}
deepfake detection, digital forensic, digital media authenticity, empirical evaluation
\end{IEEEkeywords}

\section{Introduction} \label{sec:introduction}
The proliferation of synthetic media generated through artificial intelligence (AI) has emerged as one of the most consequential challenges confronting digital societies. Deepfakes (DFs) encompass AI-generated or AI-manipulated images, video, and audio that can be nearly indistinguishable from authentic content \cite{ref1_westerlund2019}. Once confined to academic laboratories and well-resourced institutions, DF generation has become accessible to the general public through user-friendly platforms such as Google Gemini\footnote{\url{https://gemini.google.com}}, HeyGen\footnote{\url{https://www.heygen.com}}, and FaceSwap\footnote{\url{https://faceswap.dev}}, effectively democratising the production of highly realistic fabricated media. Caldwell et al. \cite{ref2_caldwell2020} ranked deepfake-enabled crime as one of the most serious AI threats for the next fifteen years, citing both its difficulty to detect and the breadth of its potential for misuse.

The societal consequences are already tangible and severe. In the personal sphere, fabricated explicit images of public figures such as Taylor Swift have spread rapidly across social media, sparking legislative debate \cite{ref9_taylor_swift}. n the domain of financial fraud, a Hong Kong-based finance officer was deceived into transferring US$\$25$ million after participating in a DF video conference impersonating his CFO and colleagues \cite{ref8_hk_deepfake_fraud}, while Deloitte estimates that losses from fraudulent misuse of generative AI will reach US$\$40$ billion in the United States alone by 2027 \cite{ref13_deloitte_genai_fraud}. In the political domain, DFs have been deployed to undermine democratic processes, including a fabricated video of President Zelensky urging Ukrainian surrender \cite{ref11_zelensky_deepfake} and AI-generated robocalls mimicking then-President Biden to mislead voters \cite{ref12_biden_robocall}. These cases illustrate that a single piece of fabricated content can upend lives, drain bank accounts, and undermine entire democracies.

In response, academia and industry have released several detection algorithms and tools. The academic community has produced a rich body of detection methods from forensic-based methods \cite{qiao2023csc, wang2022m2tr}, which explore physical and biological inconsistencies; classification-based method \cite{wang2022lisiam, tantaru2024weakly}, which distinguish real or fake samples through training with a large-scale dataset; pixel-level segmentation methods \cite{liu2023fedforgery, shi2023discrepancy}, which find manipulated regions via segmentation; and fingerprint-based methods \cite{wang2023dire, liu2020global}, which identify artifacts left by the generation process (e.g., checkerboard patterns). In parallel, a growing number of publicly accessible detection tools have emerged, ranging from forensic analysis platforms such as FotoForensics\footnote{FotoForensics \url{https://fotoforensics.com/}}, Forensically\footnote{Forensically \url{https://29a.ch/photo-forensics/\#forensic-magnifier}}, and InVID \& WeVerify\footnote{InVID \& WeVerify \url{https://weverify.eu/verification-plugin/}}, to automated AI classifiers like DecopyAI\footnote{DecopyAI  \url{https://decopy.ai/}}, FaceOnLive\footnote{FaceOnLive \url{https://faceonlive.com/deepfake-detector/}}, and Bitmind\footnote{FaceOnLive \url{https://thedetector.ai/}}. Despite these advances, a persistent gap remains between the sophistication of detection research and the practical tools that end users, particularly law enforcement investigators, rely upon in real-world scenarios.

\subsection{Motivation} \label{ssec:motivation}
There are four research gaps that motivate this work. First, existing literature \cite{nguyen2025survey, mubarak2023survey, liu2024evolving, nguyen2025deepfake, wang2024deepfake} only evaluates detection algorithms in controlled laboratory settings, while the publicly accessible tools that practitioners actually use rarely receive systematic comparison. Second, no unified evaluation framework exists for comparing across tool paradigms: forensic analysis platforms that require expert interpretation of analytical outputs (e.g., ELA maps) versus automated AI classifiers that return binary predictions and confidence scores. Third, current evaluations focus on algorithmic metrics (accuracy, precision, recall) while neglecting user-centric factors, including interpretability, transparency of results, explanatory feedback, and usability, which are equally critical for investigative and legal contexts. Fourth, there is insufficient research into human–AI discordance: when and why human judgment contradicts automated detection, and which image characteristics drive such disagreements.

This paper addresses these gaps through \textbf{a unified evaluation} of six public available DF detection tools, including three forensic analysis platforms and three automated AI classifiers, across two independent experiments with a combined 250 images, conducted by two professional investigators experienced in image and video analysis for law enforcement.

\begin{table*}[ht]
\centering
\caption{Positioning of this work against existing survey literature. \checkmark\ = addressed; \texttimes\ = not addressed; $\circ$ = partially addressed.}
\label{tab:positioning}
\resizebox{\textwidth}{!}{%
\begin{tabular}{l c c c c c c c}
\hline
\textbf{Dimension} & \textbf{Mubarak} & \textbf{Liu} & \textbf{Nguyen} & \textbf{Wang} & \textbf{Nguyen} & \textbf{This} \\
 & \textbf{et al.} \cite{mubarak2023survey} & \textbf{et al.} \cite{liu2024evolving} & \textbf{et al.} \cite{nguyen2025survey} & \textbf{et al.} \cite{wang2024deepfake} & \textbf{et al.} \cite{nguyen2025deepfake} & \textbf{work} \\
\hline
Detection algorithm taxonomy       & \checkmark & \checkmark & \checkmark & \checkmark & \checkmark & $\circ$ \\
Algorithmic benchmark (accuracy, AUC, etc.) & \checkmark & \checkmark & \checkmark & \checkmark & \checkmark & \checkmark \\
\hline
\textit{Empirical evaluation of public tools} & \texttimes & \texttimes & \texttimes & \texttimes & \texttimes & \checkmark \\
\textit{Cross-paradigm comparison (forensic platforms vs.\ AI classifiers)} & \texttimes & \texttimes & \texttimes & \texttimes & \texttimes & \checkmark \\
\textit{User-centric evaluation (interpretability, usability, transparency)} & \texttimes & \texttimes & \texttimes & \texttimes & \texttimes & \checkmark \\
\textit{Human--AI discordance analysis} & \texttimes & \texttimes & \texttimes & \texttimes & \texttimes & \checkmark \\
\textit{Practitioner-led assessment (law enforcement perspective)} & \texttimes & \texttimes & \texttimes & \texttimes & \texttimes & \checkmark \\
\textit{Confidence score \& feedback quality analysis} & \texttimes & \texttimes & \texttimes & \texttimes & \texttimes & \checkmark \\
\hline
Number of tools evaluated          & 0 & 0 & 0 & 0 & 0 & 6 \\
\hline
\end{tabular}%
}
\end{table*}

\subsection{Related Literature Review}
Existing surveys focuses on reviewing and categorize DF detection algorithms. Mubarar et al. \cite{mubarak2023survey} surveyed deep learning (DL)-based detection methods for image, video, audio and text, covering spatio-temporal analysis, GAN fingerprinting, and audio-visual consistency checks. Liu et al. \cite{liu2024evolving} showed a significant development from single- to multi-modal detection methods. Nguyen et al. \cite{nguyen2025survey} provided a comprehensive survey of proactive DF defense strategies, specifically focusing on disruption and watermarking techniques across visual and audio modalities. It systematically analyzes these approaches through various evaluation metrics and identifies current threat models and attack vectors. Other survey papers evaluate detection algorithms on the generalization and robustness perspectives, showing that existing methods cannot generalize to unseen generators and not to be robust to adversarial attacks \cite{wang2024deepfake, nguyen2025deepfake}. 

However, all focus on the efficacy of individual algorithms rather than the integrated tools practitioners employ. As shown in Table~\ref{tab:positioning}, every existing survey restricts its scope to algorithmic taxonomies, benchmark metrics, or defense strategies, leaving the real-world performance, usability, and interpretability of publicly accessible detection platforms entirely unexamined. This body of work therefore provides a rich theoretical foundation but stops short of the empirical, tool-level, and human-centred evaluation that practitioners/investigators require. The specific research gaps arising from this limitation are detailed in Section~\ref{ssec:motivation}.

\subsection{Contributions}
This paper makes the following contributions:

\begin{enumerate}
    \item \textbf{First unified, cross-paradigm benchmark of public DF detection tools.} We evaluate six tools spanning two paradigms: three forensic analysis platforms (InVID \& WeVerify, FotoForensics, Forensically) and three automated AI classifiers (DecopyAI, FaceOnLive, Bitmind), across a combined dataset of 250 images encompassing multiple generation techniques.

    \item \textbf{Multi-dimensional qualitative assessment.} Beyond accuracy, precision, recall, and F1-score, we assess interpretability, feedback transparency, anhd usability, all of which are critical for adoption in investigative and legal contexts.

    \item \textbf{Systematic human-AI discordance analysis.} We quantify agreement between expert judgment and automated outputs using Cohen's Kappa and map disagreement cases to specific image characteristics and generative techniques, revealing complementary strengths and blind spots that inform hybrid human-tool workflows.

    \item \textbf{Actionable recommendations for tool development and law enforcement.} We identify consistent failure modes, providing targeted guidance for developers and practitioners.
\end{enumerate}

\section{Background}\label{sec:background}

\begin{figure}[ht] 
    \centering
    \includegraphics[width=1\columnwidth]{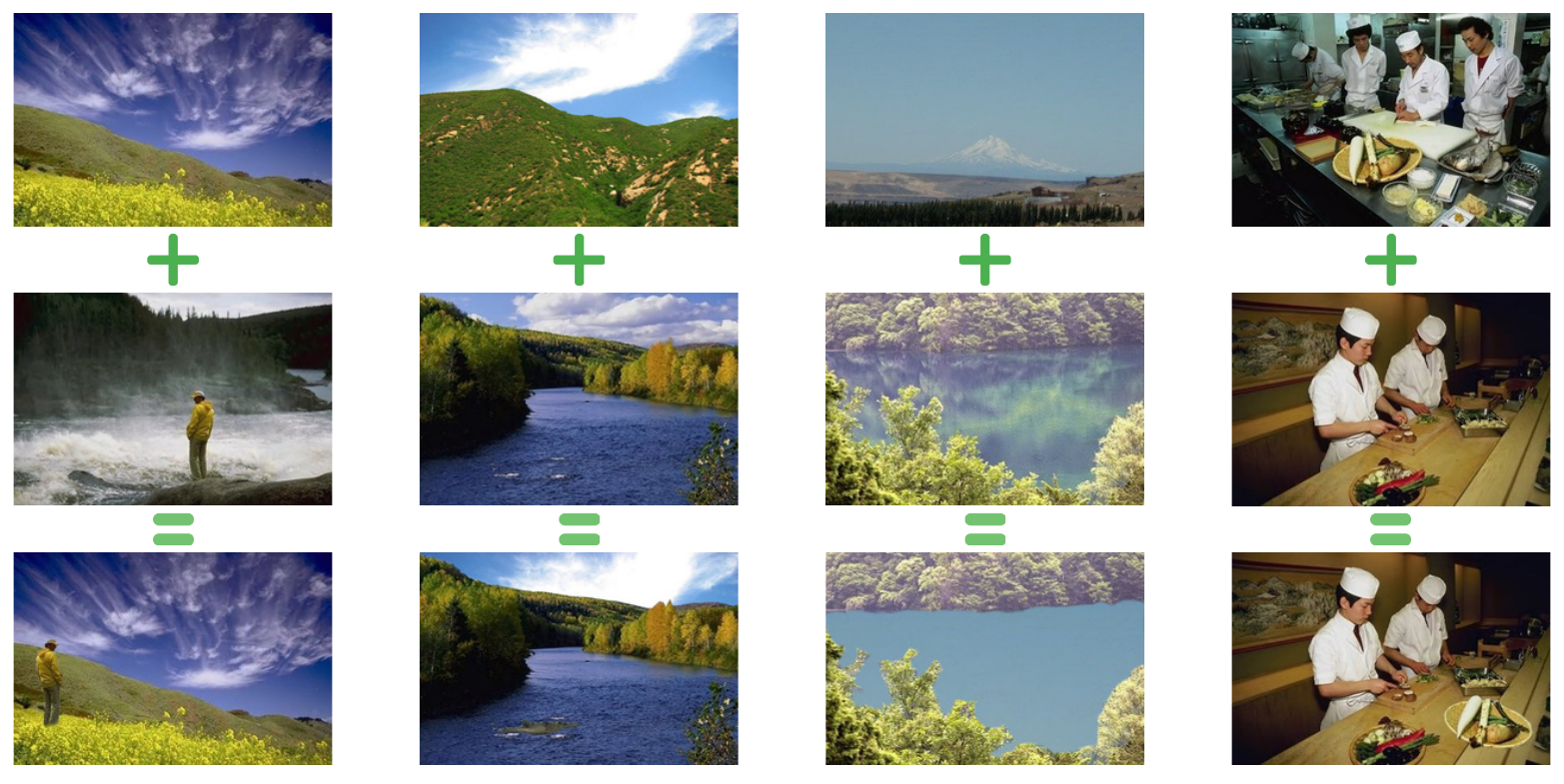}
    \caption{Example of a tampered image created via copy-move manipulation.}
    \label{fig:tampered}
\end{figure}

\begin{figure}[ht] 
    \centering
    \includegraphics[width=1\columnwidth]{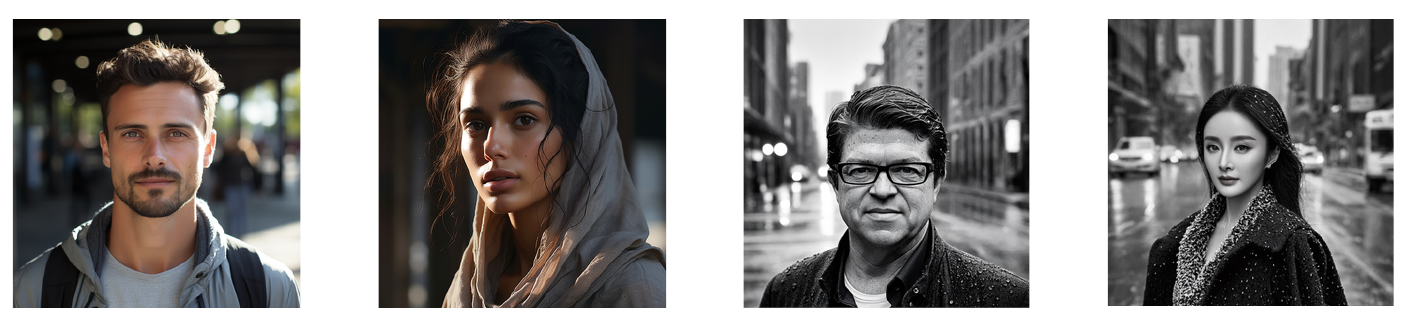}
    \caption{Example of a deepfake image generated using a diffusion model.}
    \label{fig:deepfake}
\end{figure}

Understanding the distinction between tampered and DF images is essential for interpreting detection tool performance, as each category presents fundamentally different challenges to both human analysts and automated systems. Examples of these tampered and DF images are provided in Figures \ref{fig:tampered}-\ref{fig:deepfake}.

\subsection{Tampered Images}

Image tampering refers to the deliberate modification of authentic photographs through localised manipulations. Common techniques include copy-move (duplicating a region within the same image), splicing (compositing elements from different images), and inpainting (removing or replacing objects). These operations leave detectable forensic traces, such as inconsistent noise patterns, duplicated compression artifacts, mismatched lighting, and boundary discontinuities between manipulated and original regions, because the tampered area originates from a different source or processing pipeline than the surrounding content. Traditional forensic methods, including Error Level Analysis (ELA), noise analysis, and JPEG quantisation table inspection, were specifically designed to exploit these traces and have demonstrated reasonable effectiveness against conventional tampering~\cite{da2020critical, chennamma2023comprehensive}.

\subsection{Deepfake Images}

DF images are synthetically generated or transformed using generative models (GMs), including Generative Adversarial Networks (GANs), Diffusion Models (DMs), and hybrid architectures. Unlike tampered images, DFs do not splice foreign content into an authentic photograph; instead, they synthesise entirely new pixel data or seamlessly transform existing features (e.g., face swapping, attribute editing, full scene generation). 

This fundamental difference makes DFs considerably more challenging to detect for several reasons. First, modern GMs produce statistically coherent outputs that preserve consistent noise distributions, lighting, and compression characteristics across the entire image, eliminating the localised inconsistencies that traditional forensic tools rely upon \cite{cao2024survey}. Second, the rapid proliferation of generation architectures, from StyleGAN and StarGAN to DDPM and HeyGen or MidJourney, means that detection tools must generalise across an ever-expanding diversity of synthetic signatures \cite{nguyen2025survey}. Third, DFs increasingly resist frequency-domain and pixel-level analysis as GMs incorporate adversarial training objectives that explicitly minimise detectable artifacts \cite{zhang2019detecting}. Consequently, while a tampered image typically contains a forensic artifacts between authentic and manipulated regions, a well-crafted DF presents a uniformly synthetic surface that offers no obvious point of entry for traditional detection methods.

\section{Problem Statement}\label{sec:problem}
Despite substantial progress in DF detection research, the literature remains focused on evaluating individual algorithms under controlled conditions, with little attention to the publicly accessible tools that practitioners actually employ. This leaves investigators and everyday users without evidence-based guidance on which freely available tools can reliably support DF detection in practice. To address this gap, we formulate three research questions.

\textbf{RQ1: How effective are Forensically, FotoForensics, InVID \& WeVerify, DecopyAI, FaceOnLive, and Bitmind at detecting DF and tampered images?} We hypothesise that all tools will exceed chance-level accuracy but vary substantially across generation techniques. Forensic analysis tools are expected to achieve stronger discrimination when operated by experienced analysts, at the cost of higher analysis time, while automated classifiers will deliver faster but less consistent results.

\textbf{RQ2: When and why do human evaluators and automated tools disagree, and what image characteristics drive such discordances?} We hypothesise that disagreement will peak in two scenarios: (i) high-quality DFs with subtle artifacts recognisable to human experts but missed by algorithms, and (ii) images with benign visual anomalies (compression, colour grading, motion blur) that trigger false positives in automated tools. Facial features and scene-level inconsistencies are expected to be the primary characteristics driving discordance.

\textbf{RQ3: What are the practical challenges, limitations, and usability considerations affecting the real-world utility of these tools?} We hypothesise a fundamental trade-off between analytical depth and usability: tools with greater customisation will demand more expertise and time, while simpler tools will score higher on usability but offer less diagnostic granularity. No tool is expected to reliably identify specific tampering techniques or provide the confidence-ranked, interpretable output required for evidentiary decision-making in legal contexts.

\section{Methodology}\label{sec:method}

This section describes the experimental design used to evaluate six publicly accessible DF detection tools across two paradigms. We detail the tool selection taxonomy (Section \ref{ssec:tool_overview}), dataset construction (Section \ref{ssec:datasets}), and evaluation protocols including blinded human baseline assessment and performance metrics (Section \ref{ssec:protocol}).

\subsection{Tool Selection and Overview} \label{ssec:tool_overview}

\subsubsection{Detection Tool Taxonomy}
Existing open-source DF detection tools available to end users can be broadly categorised into two fundamentally distinct paradigms based on their operational philosophy, the type of expertise they demand, and the nature of their outputs.

The first paradigm, \textbf{forensic analysis platforms}, employs signal-level techniques, such as Error Level Analysis (ELA), noise decomposition, JPEG quantisation inspection, and frequency-domain transforms, to expose statistical inconsistencies introduced by manipulation or generation. These tools do not provide an exact decision; instead, they produce visual overlays, heatmaps, and analytical layers that a trained investigator/examiner must interpret, cross-reference, and synthesise into a judgment. Their strength lies in transparency: the analytical process is fully visible, enabling the investigator/examiner to reason about \textit{why} an image may be manipulated and to present that reasoning in investigative or legal settings. However, this transparency comes at the cost of requiring domain expertise and substantially more analysis time per image.

\begin{table*}[t]
    \centering
    \caption{Summary of key characteristics for forensic and AI classifier tools.}
    \label{tab:tool-comparison}
    \small
    \newcolumntype{L}{>{\raggedright\arraybackslash}X}
    
    \begin{tabularx}{\textwidth}{l l L L p{1.8cm}}
    \toprule
    \textbf{Tool} & \textbf{Paradigm} & \textbf{Detection Approach} & \textbf{Output Format} & \textbf{User Tuning} \\ \midrule
    FotoForensics & Forensic Analysis & ELA, hidden pixels, metadata & Visual overlays requiring interpretation & None \\ \addlinespace
    Forensically & Forensic Analysis & ELA, clone detection, noise analysis, luminance grading, PCA, JPEG analysis, C2PA & Visual overlays with adjustable parameters & Extensive (per-method sensitivity sliders) \\ \addlinespace
    InVID \& WeVerify & Forensic Analysis & ELA, WAVELET, DCT, Double Quantization, CAGI, BLOCK, CMFD, Mantranet, FUSION & Visual overlays across multiple algorithm tabs & None \\ \midrule
    DecopyAI & AI Classifier & Deep learning (undisclosed) & Binary prediction + confidence score (\%) & None \\ \addlinespace
    FaceOnLive & AI Classifier & Deep learning (undisclosed) & Binary prediction + confidence score (\%) & None \\ \addlinespace
    Bitmind & AI Classifier & Deep learning (undisclosed) & Binary prediction + confidence in prediction (\%) & None \\ \bottomrule
    \end{tabularx}
\end{table*}

The second paradigm, \textbf{automated AI classifiers}, relies on DL architectures trained on large-scale datasets of real and synthetic images. These tools accept an image as input and return a binary prediction (real or fake) accompanied by a confidence score, requiring no specialised forensic knowledge from the user. Their strength is accessibility and speed: any user can obtain a result within seconds. The trade-off is opacity because these tools operate as black boxes whose underlying architectures and training data are typically undisclosed, providing no interpretable evidence to support their prediction.

This two-paradigm distinction is not merely taxonomic; it reflects a fundamental trade-off in how detection tools balance \textbf{analytical depth against accessibility}, and \textbf{transparency against automation}. Neither paradigm alone is sufficient for real-world investigative needs. Forensic analysis tools offer the interpretability and evidentiary traceability that legal contexts demand but are impractical for high-volume screening. Automated classifiers enable rapid triage but cannot explain their decisions, limiting their utility as standalone evidentiary instruments. Therefore, this study should span both paradigms to (i) benchmark their respective detection performance, (ii) expose their complementary strengths and blind spots, and (iii) inform the design of hybrid human–tool workflows that combine automated screening with expert forensic validation, which is the operational model most likely to serve investigators in practice.

\subsubsection{Selected Tools}
Three criteria guided tool selection across both paradigms: (1) public accessibility without cost barriers; (2) applicability to static image analysis; and (3) sufficient diversity in detection methodology. Specifically, for forensic analysis platforms, we select 3 popular tools: FotoForensics, Forensically, and InVID \& WeVerify; while DecopyAI, FaceOnLive, and Bitmind are selected for automated AI classifiers. Table \ref{tab:tool-comparison} summarizes characteristics of these tools and detailed information about these tools are provided in Appendix. Furthermore, all evaluations were conducted by two investigators with professional experience in image and video analysis for law enforcement purposes, ensuring that tool outputs were interpreted with operational realism rather than purely academic abstraction. 

\subsection{Datasets} \label{ssec:datasets}

The first dataset, used to evaluate the forensic analysis tools, comprises 100 images (63 fake, 37 real) sourced from three established benchmarks: DF40 \cite{yan2024df40}, CelebDF \cite{li2020celeb}, and CASIA-v2 \cite{dong2013casia}. The fake images include both fully AI-generated imagery and tampered images, spanning seven forgery types: copy objects, replace objects, remove objects, face editing, entirely generated, impersonation, and artist's style mimicry. The real images are both unedited and edited photographs to evaluate the robustness of these tools. 

The second dataset, used to evaluate the automated AI classifiers, comprises 150 images divided into facial images (70 DF, 40 real) and complete scenes depicting criminal activity (20 DF, 20 real), reflecting two prevalent DF misuse scenarios. Real images were sourced from CelebA \cite{zhu2022celebv}, FFHQ \cite{karras2019style}, VFHQ \cite{xie2022vfhq}, and UCF \cite{sultani2018real}. Fake images were generated using DMs (MidJourney-v6, DDPM, RDDM, CollabDiff), GAN-based methods (StyleGAN3, StarGAN2, other GAN), face-swapping (SimSwap), and commercial avatar generation (HeyGen).

\begin{table}[ht] 
    \centering
    \caption{Summarization of both datasets.}
    \label{tab:dataset-overview}
    \small 
    \newcolumntype{L}{>{\raggedright\arraybackslash}X}
    
    \begin{tabularx}{\columnwidth}{@{}p{1.75cm} L L@{}}
    \toprule
    \textbf{Characteristic} & \textbf{Dataset 1 (Forensic Tools)} & \textbf{Dataset 2 (AI Classifiers)} \\ \midrule
    Total images & 100 & 150 \\ \addlinespace
    Fake / Real & 63 / 37 & 90 / 60 \\ \addlinespace
    Content categories & Mixed (predominantly facial) & Facial (110) and Scene (40) \\ \addlinespace
    Real image sources & DF40, CelebDF, CASIA-v2 & CelebA, FFHQ, VFHQ, UCF \\ \addlinespace
    Generative methods & Various (via benchmark datasets) & StyleGAN3, StarGAN2, Other GAN, HeyGen, SimSwap, Diffusion Models \\ \addlinespace
    Blinding procedure & Neutral folder, no metadata cues & Randomized file naming \\ \bottomrule
    \end{tabularx}
\end{table}

\subsection{Evaluation Protocols}\label{ssec:protocol}
Both experiments were conducted by two professional investigators with extensive experience in image and video analysis for law enforcement purposes. To ensure ecological validity, the evaluation protocols were designed to simulate realistic \textbf{investigative workflows} while maintaining methodological rigor through blinded assessment and standardized procedures.

\subsubsection{Forensic Analysis Platforms Evaluation} Each of the 100 images was submitted individually to Forensically, FotoForensics, and InVID \& WeVerify. Where a tool offered multiple detection algorithms (e.g., ELA, noise analysis, clone detection, double quantization, WAVELET), all available capabilities were employed and their outputs considered jointly, consistent with each platform's recommended usage guidelines. For Forensically, whose methods offer adjustable sensitivity parameters (e.g., minimal similarity, cluster size, block size), optimal settings were identified per image type through preliminary calibration and applied consistently to balance detection granularity with reproducibility. No user-tunable parameters exist for FotoForensics or InVID \& WeVerify. 

Since none of the forensic tools provides a binary real/fake prediction, the investigator synthesised outputs across all available algorithms within each tool to arrive at a final predicted label. To mitigate interpretation bias, the full dataset was submitted to each tool before any results were compared against ground truth. Four classification metrics were computed per tool: accuracy, precision, recall, and F1-score, calculated both overall and disaggregated by image category.


\subsubsection{Automated AI Classifiers Evaluation} Each of the 150 images was submitted sequentially to DecopyAI, FaceOnLive, and Bitmind, with one image processed at a time. Each tool returned a binary prediction (real or fake) and a confidence score. Since DecopyAI and FaceOnLive report the estimated probability of manipulation (higher = more likely fake), whereas Bitmind reports confidence in its binary prediction (a high score may indicate strong certainty that an image is real), Bitmind's scores for images assessed as real were inverted so that all confidence values uniformly represent estimated deepfake probability across tools. 


\subsubsection{Human Baseline Evaluation} A blinded human evaluation was performed prior to tool assessment to establish a human baseline and enable subsequent human-AI discordance analysis. The investigator, who was blind to the ground truth and had no prior experience with any of the three automated tools, examined each image manually, using only pinch-and-zoom functionality without any filtering, color adjustment, or other digital enhancement. For each image, the evaluator recorded a binary classification (real or fake), a confidence rating on a 1--5 Likert scale (where higher values indicate greater confidence that the image is a deepfake), and the specific image characteristics (e.g., inconsistencies in eyes, skin, transitions, objects) that informed the decision. 

\subsubsection{Metrics} Four classification metrics were computed per tool: accuracy, precision, recall, and F1-score, calculated both overall and disaggregated by content category (facial vs. scene) and by generative technique. In addition to standard classification metrics, we report the Real Detection Rate (Real Det.), defined as the proportion of genuine images correctly identified as real, i.e., TN/(TN+FP). This metric captures a tool's specificity and is particularly relevant for investigative contexts where falsely accusing authentic evidence of being fabricated carries serious consequences.

\section{Results} \label{sec:results}
This section presents the empirical findings organised around the three research questions. We report overall detection performance across all six tools and the human baseline (Section \ref{ssec:results_accuracy}), disaggregate detection rates by generative technique (Section \ref{ssec:results_technique}), analyse human-AI agreement and disagreement patterns (Section \ref{ssec:results_human}), and examine representative misclassification cases for both paradigms (Section \ref{ssec:misclassification}).

\subsection{Overall Detection Accuracy}
\label{ssec:results_accuracy}

Table~\ref{tab:overall_accuracy} presents the unified performance comparison across all six tools and the human baseline. To enable direct cross-paradigm comparison, we compute accuracy, precision, recall, F1-score, real detection rate for the forensic analysis tools using their match/mismatch counts against ground truth. 

\begin{table}[t]
    \centering
    \caption{Overall detection performance of all six tools and the human evaluator. Forensic tools were evaluated on Dataset~1 (100 images); AI classifiers and the human evaluator on Dataset~2 (150 images).}
    \label{tab:overall_accuracy}
    \resizebox{\columnwidth}{!}{%
    \begin{tabular}{llccccc}
    \hline
    \textbf{Tool} & \textbf{Paradigm} & \textbf{Acc.} & \textbf{Prec.} & \textbf{Rec.} & \textbf{F1} & \textbf{Real Det.} \\
    \hline
    Forensically    & Forensic  & 0.78 & 0.83 & 0.83 & 0.83 & 0.70 \\
    FotoForensics   & Forensic  & 0.74 & 0.77 & 0.84 & 0.80 & 0.57 \\
    InVID \& WeV.   & Forensic  & 0.69 & 0.71 & 0.87 & 0.78 & 0.38 \\
    \hline
    FaceOnLive      & AI Classif. & 0.79 & 0.93 & 0.70 & 0.80 & 0.92 \\
    Bitmind         & AI Classif. & 0.68 & 0.90 & 0.52 & 0.66 & 0.92 \\
    DecopyAI        & AI Classif. & 0.66 & 0.91 & 0.48 & 0.63 & 0.93 \\
    \hline
    Human           & --          & 0.94 & 0.96 & 0.94 & 0.95 & 0.93 \\
    \hline
    \end{tabular}%
    }
\end{table}

From Table~\ref{tab:overall_accuracy}, we can observe that:
\begin{itemize}
    \item The human evaluator substantially outperforms all six tools, achieving 94\% accuracy, which is higher than the best-performing tool, FaceOnLive, by 15\%. This result confirms that experienced investigators remain the most reliable single resource for DF detection.
    \item The forensic analysis tools exhibit high recall but poor real detection rates. For example, InVID \& WeVerify detect 83\% fake images but it correctly identifies only 37.8\% of real images, performing worse than chance. This pattern arises because the layered analytical outputs produced by these platforms, including ELA maps, noise visualisations, and compression artifact indicators, routinely flag benign characteristics such as JPEG recompression or resizing artifacts as evidence of manipulation. Forensically partially mitigates this issue through more granular sensitivity controls and clearer interpretive guidance, achieving the highest real detection rate within this paradigm at 70.3\%.
    \item The AI classifiers display the opposite trade-off. All three tools achieve real detection rates above 91\%, but their recall for fake images is markedly lower. For example, DecopyAI can detect only 48\% of fake images. 
\end{itemize}

These results reveal a fundamental complementarity between the two paradigms: forensic tools are sensitive to editting techniques applied to real images although they unlikely miss a DF, whereas AI classifiers fail to detect DFs. Neither paradigm alone delivers the combination of high recall and high specificity that investigative practice demands.

\subsection{Detection by Generation Technique}
\label{ssec:results_technique}

In this experment, we aim to evaluate the detection capabilities of these tools against specific types of generation techniques. The evaluation dimensions differ across paradigms by design. Forensic analysis tools were originally constructed to detect traditional image tampering by analysing forensic artifacts such as compression inconsistencies and splicing boundaries. Therefore, we evaluate whether these same techniques generalise to fully AI-generated DFs. AI classifiers, by contrast, are black-box systems whose training data and learned representations are unknown. Since their effectiveness is likely tied to the generative architectures seen during training, we evaluate their generalization across distinct generator families (GAN-based, diffusion-based, and hybrid methods such as HeyGen) to assess whether they generalise to unseen generation pipelines. 
Table~\ref{tab:generation} disaggregates detection rates by generation technique for the AI classifiers (Dataset~2) and summarises forgery-type detection rates for the forensic tools (Dataset~1).

\begin{table}[t]
    \centering
    \caption{Detection rate by generative technique for AI classifiers (Dataset~2) and by forgery type for forensic tools (Dataset~1).}
    \label{tab:generation}
    \resizebox{\columnwidth}{!}{%
    \begin{tabular}{lccccc}
    \hline
    \multicolumn{6}{c}{\textit{AI Classifiers (Dataset~2)}} \\
    \hline
    \textbf{Technique} & \textbf{$n$} &  \textbf{DecopyAI} & \textbf{FaceOnLive} & \textbf{Bitmind} \\
    \hline
    StyleGAN3       & 10 & 100.0\% & 90.0\%  & 40.0\% \\
    StarGAN2        & 10 & 10.0\%  & 50.0\%  & 40.0\% \\
    Other GAN       & 10 & 30.0\%  & 30.0\%  & 80.0\% \\
    HeyGen          & 10 & 0.0\%  & 0.0\%  & 0.0\% \\
    SimSwap         & 10 & 70.0\%  & 90.0\%  & 70.0\% \\
    DM (facial)     & 20 & 45.0\%  & 85.0\% & 70.0\% \\
    DM (scene)      & 20 & 65.0\% & 100.0\% & 50.0\% \\
    \hline
    \multicolumn{6}{c}{\textit{Forensic Tools (Dataset~1)}} \\
    \hline
    \textbf{Forgery Type} & \textbf{$n$} & \textbf{Forensic.} & \textbf{FotoFor.} & \textbf{InVID} & \\
    \hline
    AI-generated    & 53 & 81.1\% & 81.1\% & 81.1\% & \\
    Tampered        & 19 & 84.2\% & 78.9\% & 84.2\% & \\
    Technique ID    & 63 & 22.2\% & 20.6\% & 17.5\% & \\
    \hline
    \end{tabular}%
    }
\end{table}

\textbf{Can forensic tools generalise from tampering detection to DF detection?} All three forensic tools detect AI-generated images at 81.1\%, only marginally below their tampered image detection rates (78.9--84.2\%). These results indicate that fully synthesised DFs still produce artifacts visible through ELA and noise analysis despite lacking the splicing boundaries these tools were designed to detect. However, no tool correctly identifies the specific generation or tampering technique, meaning these platforms can flag suspicious images but cannot determine how the manipulation was performed.

\textbf{Can AI classifiers generalise across unseen generator families?} The results reveal pronounced generator-dependent variation. Most strikingly, all three classifiers fail entirely on HeyGen, exposing a critical blind spot for commercial hybrid generation tools. GAN-based detection is highly uneven across tools: DecopyAI excels at StyleGAN3 (100\%) but fails on StarGAN2 (10\%), while Bitmind achieves only 40\% for both StyleGAN2 and StyleGAN3. These results suggest each classifier has learned architecture-specific artifacts and generalises poorly to unfamiliar variants. Diffusion model outputs are detected with moderate success overall, with FaceOnLive achieving 100\% on scenes, though DecopyAI and Bitmind lag on facial diffusion images (45\% and 70\%). These results confirm that no single AI classifier provides reliable coverage across the full spectrum of contemporary generation methods.

\subsection{Human vs. Tool Performance}
\label{ssec:results_human}

A direct comparison between human judgment and automated tool output is meaningful only for the AI classifier paradigm, as forensic analysis tools form a coupled human-tool system where the investigator interprets intermediate analytical outputs to reach a final decision. AI classifiers, by contrast, deliver autonomous binary predictions, enabling independent discordance analysis.

\begin{table}[t]
    \centering
    \caption{Human--AI agreement and disagreement statistics for the three AI classifiers evaluated against a blinded human baseline on 150 images.}
    \label{tab:human-ai}
    \resizebox{\columnwidth}{!}{%
    \begin{tabular}{llccc}
    \toprule
     & & \textit{DecopyAI} & \textit{FaceOnLive} & \textit{Bitmind} \\
    \midrule
    \multirow{5}{*}{\textit{\shortstack[l]{On\\Agreement}}}
     & Overall & 64\% & 77\% & 66\% \\
     & Considered DF & 43\% & 53\% & 45\% \\
     & Considered Real & 57\% & 47\% & 55\% \\
     & Correct predictions & 97\% & 97\% & 97\% \\
     & False predictions & 3\% & 3\% & 3\% \\
    \midrule
    \multirow{6}{*}{\textit{\shortstack[l]{On\\Disagreement}}}
     & Overall & 36\% & 23\% & 34\% \\
     & Human: DF / Tool: Real & 89\% & 80\% & 86\% \\
     & Human: Real / Tool: DF & 11\% & 20\% & 14\% \\
     & Human correct & 89\% & 83\% & 88\% \\
     & Tool correct & 11\% & 17\% & 12\% \\
    \midrule
    \textit{\shortstack[l]{Cohen's\\Kappa}}
     & & 0.33 & 0.54 & 0.36 \\
    \bottomrule
    \end{tabular}%
    }
\end{table}

The evaluator classified all 150 images in a blinded session prior to any tool evaluation, using only visual inspection with zoom capability. As shown in Table \ref{tab:human-ai}, when when human and tool predictions coincide, the joint decision is correct 97\% of the time across all three classifiers. This consistency holds regardless of whether the agreed label is DF or real, indicating that concordance between an experienced investigator and an AI classifier can serve as a strong confidence signal in operational workflows. However, the disagreement pattern is markedly asymmetric: the dominant discordance type is the human correctly labeling a deepfake that the tool misses (80--89\% of disagreements). This indicates that current AI classifiers remain prone to false negatives on images that an experienced investigator can identify through semantic and perceptual cues such as anatomical inconsistencies, lighting irregularities, and object-level implausibility.

In this experiment, we also use Cohen's $\kappa$ to measure agreement between human and tools, with values below 0.20 indicating slight agreement, 0.21--0.40 fair agreement, 0.41--0.60 moderate agreement, and above 0.60 substantial agreement. Under this scale, FaceOnLive ($\kappa
= 0.54$) reaches moderate agreement, yet still falls short of the substantial threshold, underscoring that even the best-performing classifier in our evaluation cannot reliably substitute for experienced human assessment. 

\begin{figure}[ht]
    \centering
    \noindent\includegraphics[width=1\columnwidth]{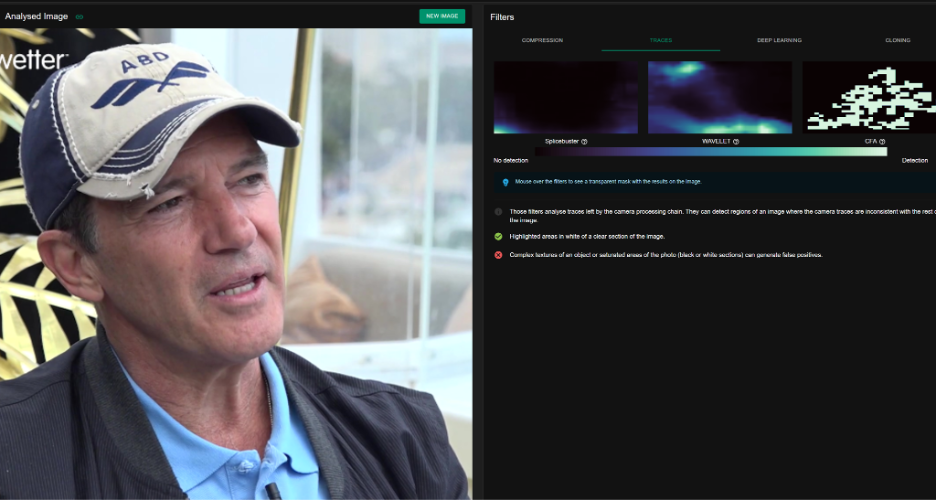}
    \vspace{0.25cm}
    \noindent\includegraphics[width=1\columnwidth]{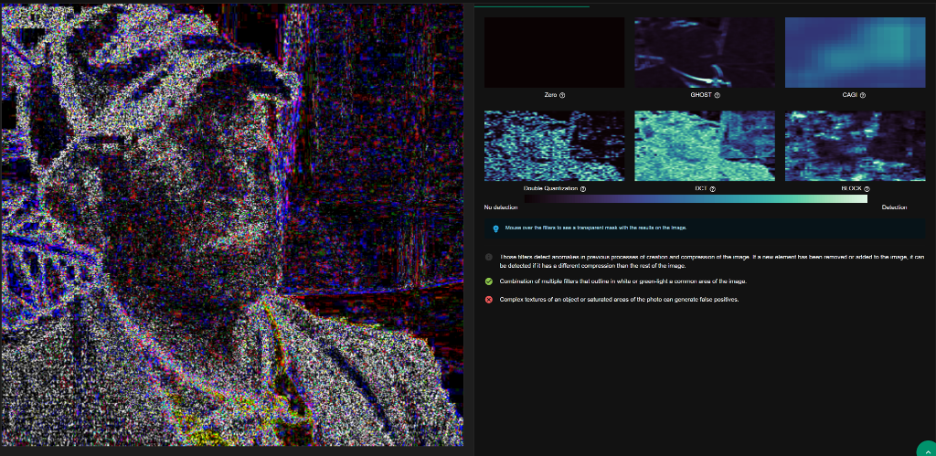}
    \caption{InVID \& WeVerify analysis of a real image. ELA and multiple compression detectors converge on false suspicion driven by natural colour contrast.}
    \label{fig:invid-fp}
\end{figure}

\subsection{Miscalssification Analysis} \label{ssec:misclassification}

\subsubsection{Forensic Analysis Tools} We illustrate representative failure modes by examining one misclassified image per tool. 

\textbf{InVID \& WeVerify (false positive).} Figure~\ref{fig:invid-fp} presents the analysis of image, a genuine, unmanipulated photograph. ELA and multiple compression detectors (Double Quantisation, WAVELET, BLOCK, GHOST) converge on overlapping regions, triggering a false manipulation prediction. The root cause is a high-saturation foreground element against a contrasting background, a benign property that activates fixed-threshold detectors with no user-adjustable sensitivity to suppress them.

\begin{figure}[ht]
    \centering
    \includegraphics[width=1\columnwidth]{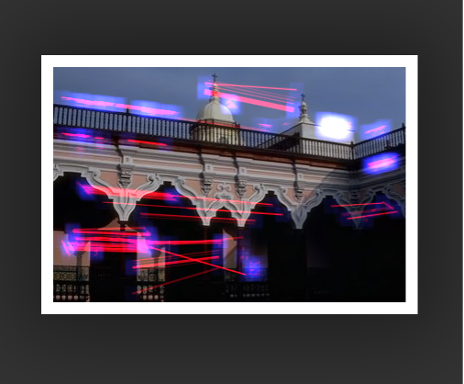}
    \caption{Forensically clone detection on a real image. Recurring architectural textures produce excessive false clone matches.}
    \label{fig:forensically-fp}
\end{figure}

\textbf{Forensically (false positive).} Figure~\ref{fig:forensically-fp} shows Forensically's clone detection output example, which is a real editted image. The tool interprets naturally recurring architectural textures as copy-move evidence, producing matches far exceeding the actual manipulation extent. Despite configurable sensitivity sliders, the structural repetitiveness of the scene defeats calibration, demonstrating that parametric control alone cannot resolve semantic ambiguity in visually repetitive imagery.

\begin{figure}[ht]
    \centering
    \begin{minipage}{0.48\columnwidth}
        \centering
        \includegraphics[width=\textwidth]{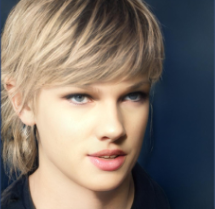}
    \end{minipage}
    \hfill 
    \begin{minipage}{0.48\columnwidth}
        \centering
        \includegraphics[width=\textwidth]{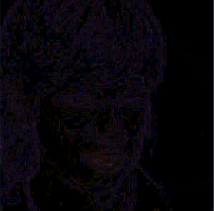}
    \end{minipage}

    \vspace{0.15cm} 
    \caption{FotoForensics ELA analysis of a fake AI-generated image. The featureless ELA map reveals no detectable manipulation traces.}
    \label{fig:fotoforensics-fn}
\end{figure}

\textbf{FotoForensics (false negative).} Figure~\ref{fig:fotoforensics-fn} displays FotoForensics' ELA output for an image example, which is a fully AI-generated face. The ELA map is effectively featureless, showing no energy differentials between regions. Because FotoForensics relies predominantly on ELA with no supplementary algorithms or tunable parameters, it has no fallback when the generative model produces uniform compression behaviour and consistent noise distributions across the entire synthesised image.

\begin{figure*}[ht]
    \centering
    \includegraphics[width=1\linewidth]{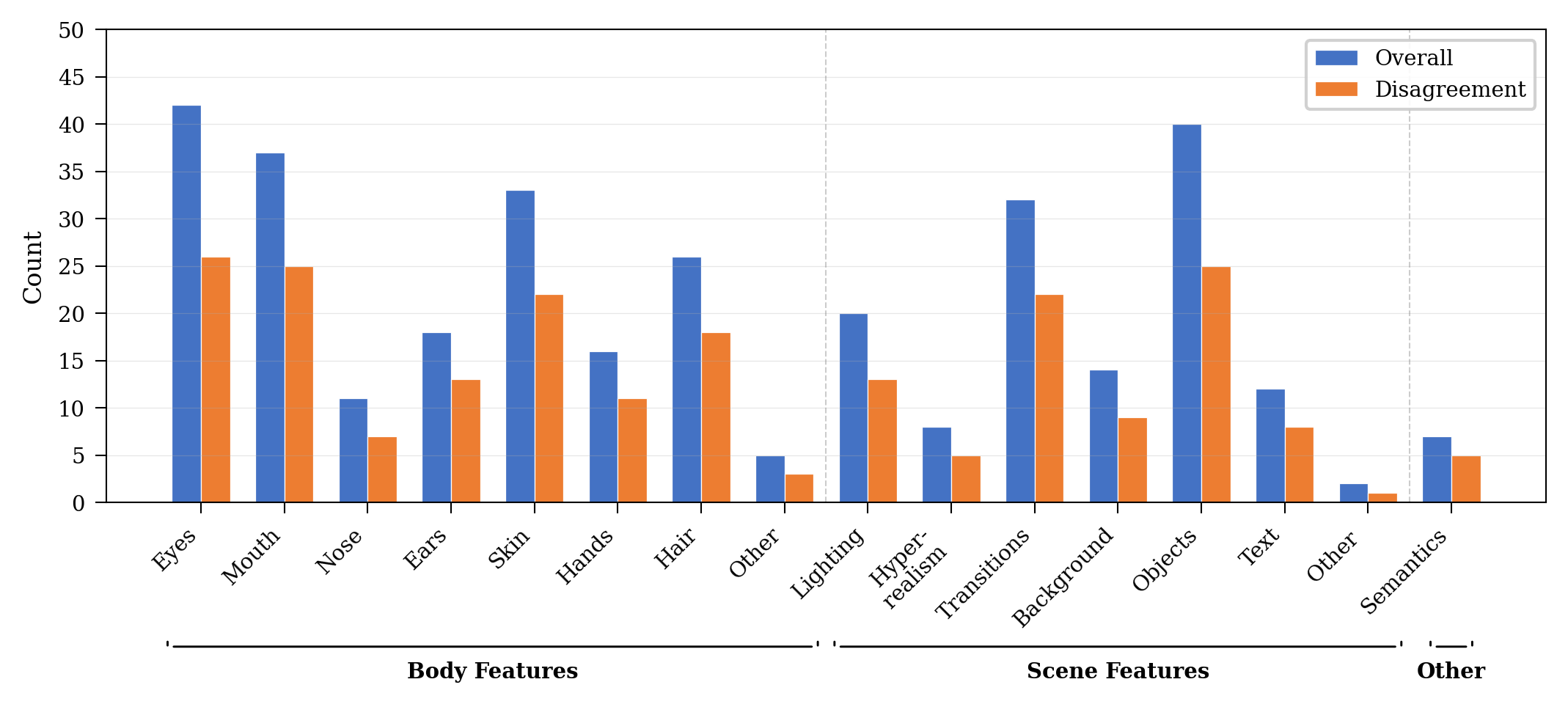}
    \caption{Distribution of image characteristics cited by the human evaluator when classifying images as DFs. Blue bars represent the full dataset; orange bars represent disagreement cases where the
    human was correct and at least one AI classifier was incorrect.}
    \label{fig:characteristics}
\end{figure*}

\subsubsection{AI Classifiers} Unlike forensic tools, whose misclassifications stem from the vulnerability of forensic-based algorithms to editting techniques, AI classifiers errors are driven by the opacity of their learned representations. To understand \emph{what} classifiers miss, we analyse the visual cues that the human evaluator used to reach correct decisions in cases where at least one tool disagreed. Figure \ref{fig:characteristics} summarises these cues across the full dataset (blue) and the disagreement cases (orange), where the human correctly identified a DF that one or more classifiers missed. To illustrate how these characteristics manifest and why they evade automated detection, we examine representative failure patterns.

\begin{figure}[htbp]
    \centering
    \noindent\includegraphics[width=1\columnwidth]{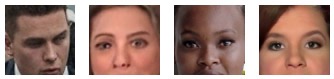}
    \vspace{0.25cm}

    \noindent\includegraphics[width=1\columnwidth]{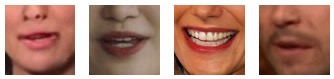}
    \caption{Eye (top) and mouth (bottom) anomalies in DFs missed by AI classifiers.}
    \label{fig:eye-mouth-anomalies}
\end{figure}

\textbf{Eyes and mouth} Eyes and mouth are the most common cues driving human-tool disagreement (Figure \ref{fig:eye-mouth-anomalies}). Eye anomalies include mismatched iris colour, asymmetric specular reflections, and unnatural edge transitions around the iris or eyelid. Mouth artifacts present as perspective distortions inconsistent with head pose, colour deviations, and teeth rendered with uniform lighting or missing inter-dental gaps. These require understanding of bilateral facial symmetry and anatomical plausibility that pixel-level classifiers do not encode.

\begin{figure}[ht]
    \centering
    \noindent\includegraphics[width=1\columnwidth]{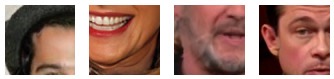}
    \vspace{0.25cm}

    \noindent\includegraphics[width=1\columnwidth]{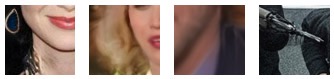}
    \caption{Skin anomalies (top) and boundary transition artifacts (bottom) in DFs missed by AI classifiers.}
    \label{fig:skin-transition-anomalies}
\end{figure}

\textbf{Skin and transitions} include localised colour shifts and anatomically implausible wrinkle textures, alongside boundary artifacts between adjacent elements (hairline-to-forehead, skin-to-clothing) that appear over-smoothed or artificially sharp (Figure~\ref{fig:skin-transition-anomalies}). These cues are perceptually salient to a trained analyst scanning for coherence across regions but are challenging to current classifiers which are trained primarily on global image statistics and tend to overlook.

\begin{figure}
    \centering
    \includegraphics[width=1\columnwidth]{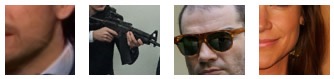}
    \caption{Object rendering errors in DFs missed by AI classifiers}
    \label{fig:objects}
\end{figure}

\textbf{Objects} Object rendering errors occur with equal frequency to mouth anomalies and are particularly prevalent in scene-depicting images. As shown in Figure~\ref{fig:objects}, generated objects frequently lack fine structural detail: weapons missing mechanical subcomponents, accessories appearing to levitate without attachment points, and normally straight edges rendered with slight curvature. 

\section{Discussion and Future Works} \label{sec:discussion-future}

This section synthesises the empirical findings into broader implications for DF detection practice. We discuss tool effectiveness and limitations (Section \ref{ssec:discuss}), provide actionable recommendations for law enforcement practitioners (Section \ref{ssec:suggest-law}), and identify priorities for future tool development (Section \ref{ssec:future}).

\subsection{Discussion} \label{ssec:discuss}
\subsubsection{Effectiveness of Current Tools}
The results in Tables \ref{tab:overall_accuracy}-\ref{tab:human-ai} confirm that no single tool delivers reliable detection across all generative methods and image categories. The cross-paradigm comparison reveals structurally different failure profiles: forensic tools achieve high recall but poor specificity, routinely flagging authentic images as manipulated, while AI classifiers show the inverse pattern with strong real detection but substantially lower recall for fake images.

\subsubsection{Challenges and Limitations}
There are three main limitations of current public DF detection tools.

First, forensic platforms analyse low-level statistical properties (noise distributions, ELA residuals, quantization tables) that are perturbed by any post-processing, whether malicious or routine. With editted images, these tools struggle to analyze correctly. In contrast, AI classifiers rely on learned synthetic-artifact representations that may be absent from unseen generator families. The complete failure of all three classifiers on HeyGen (Table \ref{tab:generation}) illustrates this: commercial hybrid pipelines can fall entirely outside learned decision boundaries.

Second, although AI classifiers can detect both tampered and full AI-generated images, they fail to identify specific generation or tampering techniques. Confidence scores can be misleading, remaining high even for false negatives, meaning practitioners cannot distinguish well-founded classifications from confident errors without independent verification. This challenges their limit practical utility. 

Third, all tools process single static images without batch processing, multi-modal support, or machine-readable output, creating significant barriers for professional workflows involving large evidence volumes.



\subsection{Suggestion for Law Enforcement and Practitioners} \label{ssec:suggest-law}

\textbf{Adopt multi-tool, multi-paradigm workflows.} Practitioners should combine forensic platforms (high sensitivity) with AI classifiers (high specificity), beginning with automated screening for rapid triage followed by forensic analysis of flagged images. 

\textbf{Treat outputs as intelligence, not predictions.} Consistent with established principles of corroboration, detection outputs should be one input among several. The high false-negative rates of classifiers and false-positive rates of forensic tools preclude reliance on any single tool.

\textbf{Document tool limitations in reporting.} Practitioners should state which tools were used, their known limitations, and how human judgment was integrated. Since tool performance changes as models are updated, results should be treated as time-stamped assessments.

\subsection{Suggestion for Future Tool Development} \label{ssec:future}

\textbf{Implement explanatory feedback.} Tools should provide visual annotations (heatmaps, noise maps, artifact highlighting), textual descriptions of anomalies, and model-level feedback showing sub-model assessments. This supports user trust and strengthens evidentiary utility.

\textbf{Standardise confidence scores and output formats.} Current confidence metrics lack cross-tool comparability. A standardised metric with consistent semantics would enable meaningful comparison. Machine-readable formats (JSON, XML) would support integration with forensic pipelines.

\textbf{Enable collaborative workflows and reporting.} Tools should support activity logging, media annotation, metadata-based filtering, and exportable reports containing predictions, confidence levels, explanatory feedback, and analyst notes.

\textbf{Expand training data diversity.} The generator-dependent variation observed in AI classifier performance indicates insufficient coverage of contemporary methods. Training datasets should incorporate commercial hybrid platforms (e.g., HeyGen), emerging architectures, and adversarial examples, with regular retraining cycles.

\section{Conclusion} \label{sec:conclusion}
This paper presented the first unified, cross-paradigm evaluation of six publicly accessible deepfake detection tools, conducted by two professional investigators with law enforcement experience across a combined 250 images. The results reveal fundamental complementarity between paradigms: forensic analysis tools offer high sensitivity at the cost of frequent false positives, while AI classifiers provide strong specificity but miss a substantial proportion of deepfakes, with critical blind spots for certain generator families. Human evaluators significantly outperform all tools, and when human and tool predictions agree, accuracy is near-perfect. These findings advocate for hybrid workflows that combine multi-tool screening with expert judgment, rather than reliance on any single tool. We further identify transparency, explainability, and standardised output as key development priorities. Future work should expand to larger datasets, additional tools including commercial platforms, video and audio modalities, multiple human evaluators, and longitudinal monitoring of tool evolution as generative techniques continue to advance.

\newpage
\small
\bibliographystyle{IEEEtran}

\bibliography{references} 

\newpage

\appendices

\section{Detailed Tool Description} \label{appendix:detailed-tool}

\begin{figure*}
    \centering
    \includegraphics[width=1\linewidth]{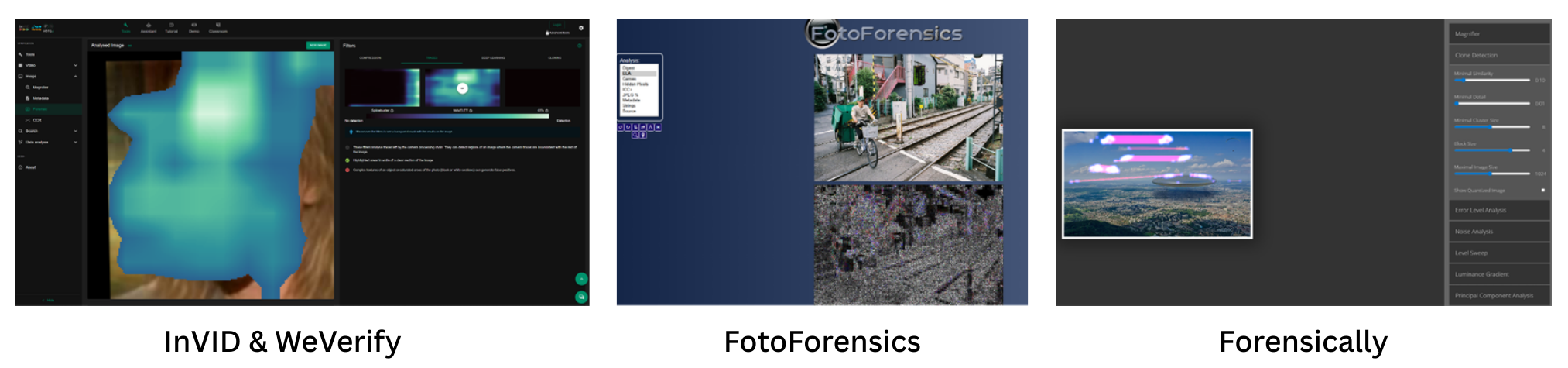}
    \caption{User interfaces of the three forensic analysis platforms. From left to right: \textbf{InVID \& WeVerify} displaying ELA and multi-algorithm forensic overlays via its browser plugin; \textbf{FotoForensics} showing its ELA output with the original image; and \textbf{Forensically} presenting its clone detection view with adjustable sensitivity parameters. These tools produce analytical visualisations that require expert interpretation rather than automated predictions.}
    \label{fig:forensic-interface}
\end{figure*}

\textbf{Forensic Analysis Platforms.} FotoForensics\footnote{FotoForensics \url{https://fotoforensics.com/}} is a web-based platform maintained by Hacker Factor, inspired by research presented at the BlackHat conference in 2007. It relies predominantly on ELA to detect manipulated or generated imagery, supplemented by metadata inspection, hidden pixel analysis, ICC profile review, and JPEG quality percentage examination. It offers no user-adjustable parameters for any of its detection capabilities, making it the most straightforward but also the most limited of the three forensic platforms. Supported input formats include JPEG, PNG, WebP, HEIC, and AVIF. Notably, FotoForensics does not guarantee upload privacy on its public-facing interface, though a dedicated login option and TOR-based submission are available.

Forensically\footnote{Forensically \url{https://29a.ch/photo-forensics/\#forensic-magnifier}} is a browser-based suite of image detection capabilities that provides substantially broader analytical depth. Its toolset includes ELA, clone detection with configurable matching parameters, noise analysis via reverse denoising, level sweep, luminance grading, principal component analysis, metadata and geotag extraction, thumbnail analysis, JPEG quantization table inspection, C2PA content authenticity verification, and string extraction. A distinguishing feature of Forensically is the granularity of user control: each detection method offers adjustable sensitivity parameters via sliding scales for variables such as minimal similarity, minimal detail, cluster size, and block size. This configurability enables fine-tuned analysis tailored to specific image characteristics, though it demands a deeper understanding of the underlying detection mechanics and significantly more time per analysis. Forensically also offers stronger privacy protections, stating that uploaded images remain on the user's local machine and are never transmitted to external servers.

InVID \& WeVerify\footnote{InVID \& WeVerify \url{https://weverify.eu/verification-plugin/}} operates as a browser plugin rather than a standalone website, offering the most extensive collection of detection algorithms among the forensic tools. Its forensic module includes ELA, noise analysis via Splicebuster and WAVELET decomposition, compression artifact detection through Double Quantization and Discrete Cosine Transform (DCT) coefficient analysis, grid-based methods (CAGI and BLOCK), copy-move forgery detection (CMFD), and experimental deep learning algorithms (Mantranet and FUSION). The platform additionally provides reverse image search integration through Google Lens, Baidu, Yandex, and Google Factcheck, as well as magnification and OCR capabilities. Unlike Forensically, InVID \& WeVerify does not permit user adjustment of detection sensitivity, instead relying on the aggregation of outputs across multiple algorithms to establish suspicion. Its own documentation acknowledges that individual detectors may flag suspicious activity in both authentic and manipulated images, and that confidence in a determination should be drawn from convergence across multiple capabilities.

\begin{figure*}
    \centering
    \includegraphics[width=1\linewidth]{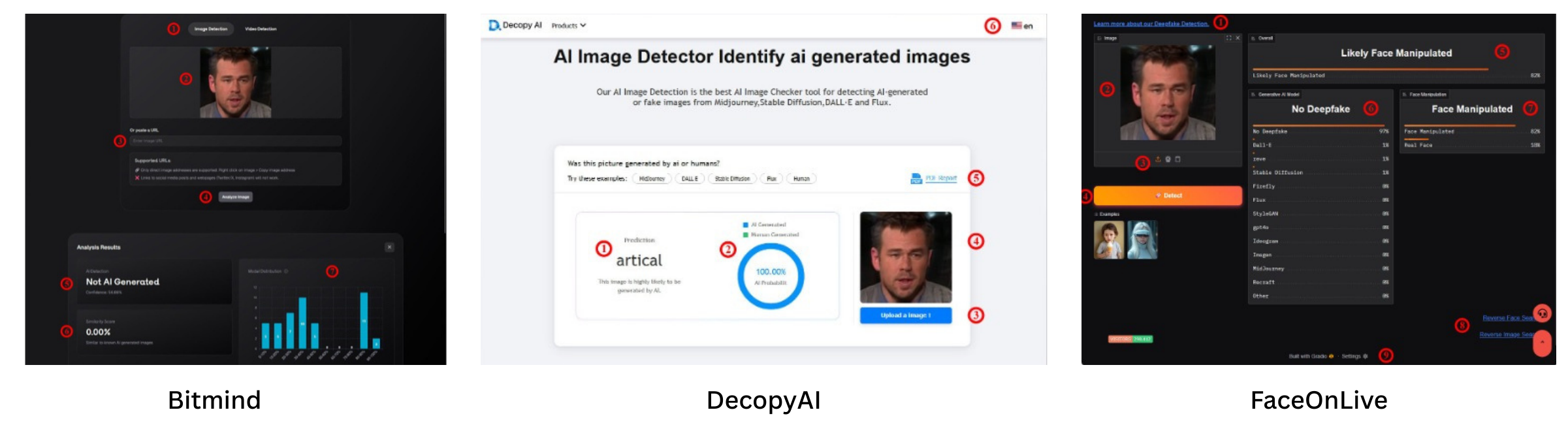}
    \caption{User interfaces of the three automated AI classifiers. From left to right: Bitmind (1: media selection; 2: image preview; 3: alternative media selection via URL; 4: "Analyze Image" button; 5: prediction and confidence score; 6: similarity score; 7: distribution of individual confidence scores across all used models); DecopyAI (1: prediction; 2: confidence score; 3: upload button; 4: image preview; 5: PDF report; 6: language settings); and FaceOnLive (1: “Learn more” button; 2: image preview; 3: source selection; 4: “Detect” button; 5: overall prediction and confidence score; 6: DF prediction and confidence score; 7: face manipulation prediction and confidence score; 8: links to other tools; 9: language settings).}
    \label{fig:placeholder}
\end{figure*}

\textbf{Automated AI Classifiers.} DecopyAI\footnote{DecopyAI  \url{https://decopy.ai/}}, FaceOnLiveFaceOnLive\footnote{FaceOnLive \url{https://faceonlive.com/deepfake-detector/}}, and Bitmind\footnote{FaceOnLive \url{https://thedetector.ai/}} are web-based DF detection services that require only a computer, an internet connection, and a standard browser. No installation, programming, specialized hardware, or provision of personal data is required for any of the three tools, making them accessible to users with minimal technical expertise. All three operate as black-box classifiers whose underlying architectures and training data are not publicly disclosed, and none offers user-adjustable detection parameters. This opacity contrasts sharply with the forensic analysis tools, where the analytical process is transparent even when interpretation remains subjective.

DecopyAI and FaceOnLive express their output as the estimated probability of manipulation: a higher score indicates greater likelihood that an image is a DF. Bitmind employs a different scoring convention, reporting the confidence level in its binary prediction rather than the probability of manipulation directly. Consequently, an image classified as real may receive a high confidence score from Bitmind, indicating strong certainty in the real assessment rather than a high probability of being fake. For cross-tool comparability, Bitmind's scores for images assessed as real must be inverted so that all confidence values uniformly represent estimated DF probability.

\section{Full Dataset Details} \label{appendix:detail-dataset}

This appendix provides comprehensive details on the two datasets used to evaluate the forensic analysis platforms and automated AI classifiers, respectively. Both datasets were constructed to prioritise generative diversity over volume, ensuring that tool performance could be assessed across a broad spectrum of manipulation and generation techniques rather than being dominated by a single synthetic pipeline. Table \ref{tab:dataset-overview} in the main paper provides a high-level summary; the following subsections expand each dataset in full.

\subsection{Dataset 1: Evaluate Forensic Analysis Tools}
\label{appendix:dataset1}
Dataset~1 comprises 100 images (63 fake, 37 real) assembled from three established benchmarks: DF40~\cite{yan2024df40}, CelebDF~\cite{li2020celeb}, and CASIA-v2~\cite{dong2013casia}. The use of multiple sources ensures that the forensic tools encounter a variety of generation algorithms representative of content found in real-world investigative scenarios.

\subsubsection{Composition by Authenticity and Forgery Type}
The 63 fake images span two broad categories and seven forgery types:

\begin{itemize}
    \item \textbf{AI-generated images} (53 images): Entirely synthesised by generative models. These include full-scene generation and face synthesis.
    \item \textbf{Tampered images} (10 images): Authentic photographs subjected to localised manipulation, including copy-move, splice, and inpainting operations.
\end{itemize}

The seven forgery types represented in the fake subset are: \textit{copy objects} (duplicating a region within or across images), \textit{replace objects} (substituting elements with foreign content), \textit{remove objects} (inpainting to erase elements), \textit{face editing} (attribute manipulation such as expression or age modification), \textit{entirely generated} (full synthetic creation with no real source), \textit{impersonation} (identity transfer via face swap or reenactment), and \textit{artist's style mimicry} (generation in the visual style of a specific artist or aesthetic). Table~\ref{tab:dataset1_forgery} details the distribution.

\begin{table}[h]
    \centering
    \caption{Dataset~1 composition by forgery type.}
    \label{tab:dataset1_forgery}
    \begin{tabular}{llr}
    \hline
    \textbf{Category} & \textbf{Forgery Type} & \textbf{Count} \\
    \hline
    \multirow{5}{*}{AI-Generated} & Entirely generated & 20 \\
     & Impersonation & 10 \\
     & Face editing & 10 \\
     & Artist's style mimicry & 3 \\
     & Other AI-generated & 10 \\
    \hline
    \multirow{3}{*}{Tampered} & Copy objects & 6 \\
     & Replace objects & 5 \\
     & Remove objects & 8 \\
    \hline
    \multirow{2}{*}{Real} & Unedited (raw) & 18 \\
     & Edited (non-malicious) & 19 \\
    \hline
    \multicolumn{2}{l}{\textbf{Total}} & \textbf{100} \\
    \hline
    \end{tabular}
\end{table}

The 37 real images include both fully unedited photographs and images that have undergone non-malicious editing (e.g., cropping, brightness adjustment, resizing). The inclusion of edited real images is to evaluate the robustness of forensic tools because, in practice, virtually all digital photographs undergo some form of post-processing during capture, storage, or transmission~\cite{da2020critical}. If the dataset contained only pristine originals, the evaluation would overestimate real-image detection rates by presenting forensic tools with an unrealistically clean baseline.



\subsection{Dataset~2: Evaluate Automated AI Classifiers}
\label{appendix:dataset2}
Dataset~2 comprises 150 static images divided into two content categories that reflect prevalent DF misuse scenarios:

\begin{itemize}
    \item \textbf{Facial images} (110 images): 70 DFs and 40 real images, representing identity-centric threats such as face swapping, attribute manipulation, and synthetic portrait generation.
    \item \textbf{Scene images} (40 images): 20 DFs and 20 real images depicting complete scenes (including criminal activity scenarios), representing the use of generative AI for disinformation and fabricated evidence.
\end{itemize}

This two-category structure enables disaggregated performance analysis, revealing whether detection tools exhibit category-specific strengths or weaknesses.

\subsubsection{Real Image Sources}

The 60 real images (40 facial, 20 scene) were sourced from four publicly available datasets:

\begin{itemize}
    \item \textbf{CelebA}~\cite{zhu2022celebv}: Large-scale face attributes dataset providing diverse celebrity facial images across a range of poses, lighting conditions, and backgrounds.
    \item \textbf{FFHQ}~\cite{karras2019style}: Flickr-Faces-HQ, a high-quality face dataset originally created for training StyleGAN, offering $1024 \times 1024$ resolution images with considerable variation in age, ethnicity, and accessories.
    \item \textbf{VFHQ}~\cite{xie2022vfhq}: Video Face Super-Resolution dataset providing high-quality facial frames extracted from video, introducing natural motion-related characteristics (e.g., slight blur, varying focus) absent from curated still-image datasets.
    \item \textbf{UCF}~\cite{sultani2018real}: UCF Crime dataset providing real-world surveillance footage frames for the scene category, ensuring authentic visual characteristics typical of security and evidentiary contexts.
\end{itemize}

\begin{table}[ht]
    \centering
    \caption{Dataset~2 composition by generative method and content category.}
    \label{tab:dataset2_gen}
    \begin{tabular}{>{\raggedright\arraybackslash}p{4cm}ccc}
    \hline
    \textbf{Generative Method} & \textbf{Facial} & \textbf{Scene} & \textbf{Total} \\
    \hline
    \multicolumn{4}{l}{\textit{GAN-based methods}} \\
    \quad StyleGAN-v3 & 10 & -- & 10 \\
    \quad StarGAN-v2 & 10 & -- & 10 \\
    \quad Other GAN & 10 & -- & 10 \\
    \hline
    \multicolumn{4}{l}{\textit{Diffusion models}} \\
    \quad MidJourney-v6, DDPM, RDDM, CollabDiff & 20 & 20 & 40 \\
    \hline
    \multicolumn{4}{l}{\textit{Face-swapping}} \\
    \quad SimSwap & 10 & -- & 10 \\
    \hline
    \multicolumn{4}{l}{\textit{Commercial hybrid}} \\
    \quad HeyGen & 10 & -- & 10 \\
    \hline
    \multicolumn{4}{l}{\textit{Real images}} \\
    \quad CelebA, FFHQ, VFHQ & 40 & -- & 40 \\
    \quad UCF & -- & 20 & 20 \\
    \hline
    \textbf{Total} & \textbf{110} & \textbf{40} & \textbf{150} \\
    \hline
    \end{tabular}
\end{table}

\subsubsection{DF Generation Methods}
The 90 DF images (70 facial, 20 scene) were generated using a diverse set of publicly available models and techniques, selected for both their popularity and ease of use. Table~\ref{tab:dataset2_gen} details the distribution across generative methods and content categories.

\begin{table}[ht]
    \centering
    \caption{Distribution of image resolutions in Dataset~2, grouped by longest dimension.}
    \label{tab:resolution}
    \begin{tabular}{lrr}
    \hline
    \textbf{Resolution Group} & \textbf{Deepfake} & \textbf{Real} \\
    \hline
    $>$ 0 pixels (smallest) & 54 & 12 \\
    $>$ 500 pixels & 13 & 21 \\
    $>$ 1000 pixels & 19 & 21 \\
    $>$ 1500 pixels & 4 & 6 \\
    \hline
    \textbf{Total} & \textbf{90} & \textbf{60} \\
    \hline
    \end{tabular}
\end{table}

\begin{table*}[ht]
    \centering
    \caption{Deepfake detection rate by image resolution for AI classifiers and the human evaluator. Metric: ACC.}
    \label{tab:resolution}
    \begin{tabular}{lcccc}
    \hline
    \textbf{Resolution} & \textbf{Human} & \textbf{DecopyAI} & \textbf{FaceOnLive} & \textbf{Bitmind} \\
    \hline
    $\leq$500\,px ($n{=}54$) & 94\% & 44\% & 52\% & 52\% \\
    501--1000\,px ($n{=}13$) & 92\% & 31\% & 100\% & 62\% \\
    1001--1500\,px ($n{=}19$) & 95\% & 68\% & 95\% & 53\% \\
    $>$1500\,px ($n{=}4$) & 100\% & 50\% & 100\% & 25\% \\
    \hline
    \end{tabular}
\end{table*}

The generative methods were chosen to span the three dominant paradigms of contemporary image synthesis:

\begin{itemize}
    \item \textbf{GAN-based methods} (30 images) include StyleGAN-v3 (unconditional face generation with alias-free synthesis), StarGAN-v2 (multi-domain image-to-image translation), and other GAN architectures, collectively representing the most established family of generative models.
    \item \textbf{Diffusion models} (40 images) include MidJourney-v6, Denoising Diffusion Probabilistic Models (DDPM), Residual Denoising Diffusion Models (RDDM), and CollabDiff, representing the rapidly maturing diffusion-based paradigm that has come to dominate high-fidelity image generation.
    \item \textbf{SimSwap} (10 images) represents dedicated face-swapping techniques that operate on existing source photographs, a distinct manipulation modality from full-image synthesis.
    \item \textbf{HeyGen} (10 images) represents commercial hybrid platforms that combine multiple generative techniques (e.g., face reenactment, lip sync, avatar creation) into consumer-facing products, creating outputs whose synthetic signatures may differ substantially from those of pure research models.
\end{itemize}

\subsubsection{Image Resolution Distribution}

The dataset exhibits considerable variation in image resolution. Table~\ref{tab:resolution} groups images by their longest dimension, split by authenticity.

The resolution distribution is not uniform across classes: DF images skew toward lower resolutions (60\% below 500 pixels on their longest dimension), while real images distribute more evenly across resolution groups. This imbalance reflects authentic conditions, as many generative models default to fixed output resolutions (e.g., $256 \times 256$ or $512 \times 512$), whereas real photographs are captured at native sensor resolution and may be subsequently downscaled. However, this distribution also has implications for detection performance. As discussed in Appendix \ref{appendix:experiments}, all three AI classifiers exhibit degraded accuracy on images below $500 \times 500$ pixels, suggesting that low-resolution inputs reduce the statistical features available for classification.

\subsection{Cross-Dataset Design Rationale}
\label{appendix:dataset_rationale}

The use of two separate datasets, rather than a single unified corpus, is a deliberate design choice motivated by the fundamentally different evaluation requirements of the two tool paradigms. Forensic analysis tools require images that test their ability to detect both traditional tampering (which leaves localised forensic traces) and AI-generated content (which produces globally consistent synthetic signatures). Dataset~1 therefore includes tampered images alongside AI-generated content, with forgery-type annotations that enable fine-grained analysis of whether tampering-detection techniques generalise to generative content.

AI classifiers, by contrast, return binary predictions without access to intermediate forensic signals, and their performance is expected to vary primarily as a function of the generative architecture used. Dataset~2 therefore prioritises generative diversity, spanning three GAN families, four diffusion model variants, a face-swapping method, and a commercial hybrid platform, while also introducing a content-category dimension (facial vs. scene) that is absent from Dataset~1.

Despite their structural differences, both datasets share key design principles: deliberate inclusion of challenging edge cases (edited real images in Dataset~1; low-resolution images and commercial hybrid outputs in Dataset~2), and sourcing from multiple established benchmarks to avoid single-dataset artefacts.

\section{Additional Experiments} \label{appendix:experiments}

\subsection{Performance by Image Resolution}
\label{sec:appendix-resolution}

To examine how input resolution affects classifier performance, we disaggregate detection rates for the 90 DFs in Dataset~2 by longest dimension. Table~\ref{tab:resolution} reports results across four resolution bins.

All three classifiers perform worst on images below 500\,px, where detection rates fall to 44--52\%, despite this bin containing the majority of DFs (60\%). This is practically significant because many generative models produce fixed-resolution outputs (e.g., 256$\times$256), and evidence images frequently arrive at reduced resolution due to platform compression or social media redistribution. Resolution sensitivity also varies across tools: FaceOnLive achieves near-perfect detection above 500\,px, while DecopyAI drops to just 31\% in the 501--1000\,px range. The human evaluator, by contrast, remains stable at 92--100\% across all bins, confirming that the semantic cues driving human detection are more resilient to spatial downsampling than the pixel-level statistical features that classifiers rely upon.

\begin{table}[ht]
    \centering
    \caption{Processing time statistics for AI classifiers across 150 images.}
    \label{tab:speed}
    \begin{tabular}{lccc}
    \hline
    \textbf{Metric} & \textbf{DecopyAI} & \textbf{FaceOnLive} & \textbf{Bitmind} \\
    \hline
    Total time (s) & 1020.38 & 408.83 & 781.20 \\
    Min / Max (s) & 0.92 / 8.43 & 1.61 / 5.71 & 3.03 / 13.96 \\
    Mean per image (s) & 6.80 & 2.73 & 5.21 \\
    Throughput (KB/s) & 54.40 & 135.79 & 71.06 \\
    \hline
    \end{tabular}
\end{table}

\subsection{Cross-Tool Speed Performance}
\label{appendix:speed}

Processing speed is a practical consideration for investigative workflows involving large evidence volumes. Table~\ref{tab:speed} summarises the timing characteristics of the three AI classifiers across the full 150-image dataset (55.52\,MB total).

FaceOnLive is the fastest tool by a substantial margin, processing images at 2.5$\times$ the speed of DecopyAI and 1.9$\times$ that of Bitmind. It also exhibits the narrowest timing variance (1.61--5.71\,s), indicating more predictable behaviour across different image sizes and content types. DecopyAI is the slowest overall, while Bitmind shows the widest per-image variance (3.03--13.96\,s).

For reference, the blinded human evaluation was completed in a continuous 1:47:33 session, yielding an average analysis time of 43.02 seconds per image. While not directly comparable to tool processing time (as human analysis includes visual inspection, reasoning, and note-taking rather than upload and computation), this figure contextualises the speed advantage of automated classifiers: even the slowest tool is approximately 6$\times$ faster per image than expert manual assessment. This speed differential becomes operationally significant when triaging large evidence sets, supporting the hybrid workflow recommended in Section \ref{sec:discussion-future} where automated classifiers perform initial screening before targeted expert review.

\subsection{Human-AI Discordance across Generation Techniques}
\label{sec:appendix-discordance-techniques}

The aggregate discordance statistics in Table~6 of the main paper conceal substantial variation across content categories and generative techniques. This section disaggregates these patterns to identify where human--tool divergence concentrates.

\subsubsection{Category-Level Discordance}

Table~\ref{tab:discordance-category} separates agreement and disagreement statistics for facial images (110) and scene images (40).

\begin{table}[h]
    \centering
    \setlength{\tabcolsep}{3pt} 
    \caption{Human--AI agreement and disagreement by category.}
    \label{tab:discordance-category}
    
    \begin{tabular}{>{\centering\arraybackslash}p{0.4\columnwidth}lccc}
    \hline
    & & \textbf{Decopy} & \textbf{FaceOn} & \textbf{Bitmind} \\
    \hline
    \multirow{2}{=}{\centering Agreement rate}
    & Facial & 57\% & 71\% & 64\% \\
    & Scene & 83\% & 93\% & 73\% \\
    \hline
    \multirow{2}{=}{\centering Correct when agreeing}
    & Facial & 95\% & 96\% & 96\% \\
    & Scene & 100\% & 100\% & 100\% \\
    \hline
    \multirow{2}{=}{\centering Human correct in disagreement}
    & Facial & 87\% & 81\% & 85\% \\
    & Scene & 100\% & 100\% & 100\% \\
    \hline
    \multirow{2}{=}{\centering Cohen's $\kappa$}
    & Facial & 0.22 & 0.44 & 0.32 \\
    & Scene & 0.65 & 0.85 & 0.45 \\
    \hline
    \end{tabular}
\end{table}

Agreement is consistently higher for scene images (73--93\%) than facial images (57--71\%), and Cohen's $\kappa$ improves correspondingly. Notably, in scene-category disagreements, human judgment is correct 100\% of the time across all three tools, and no tool outperforms the human in any disagreement case. This indicates that scene-depicting DFs, while better detected overall by both humans and tools, present fewer ambiguous boundary cases than facial images.

\subsubsection{Technique-Level Discordance}

\begin{figure*}
    \centering
    \includegraphics[width=1\linewidth]{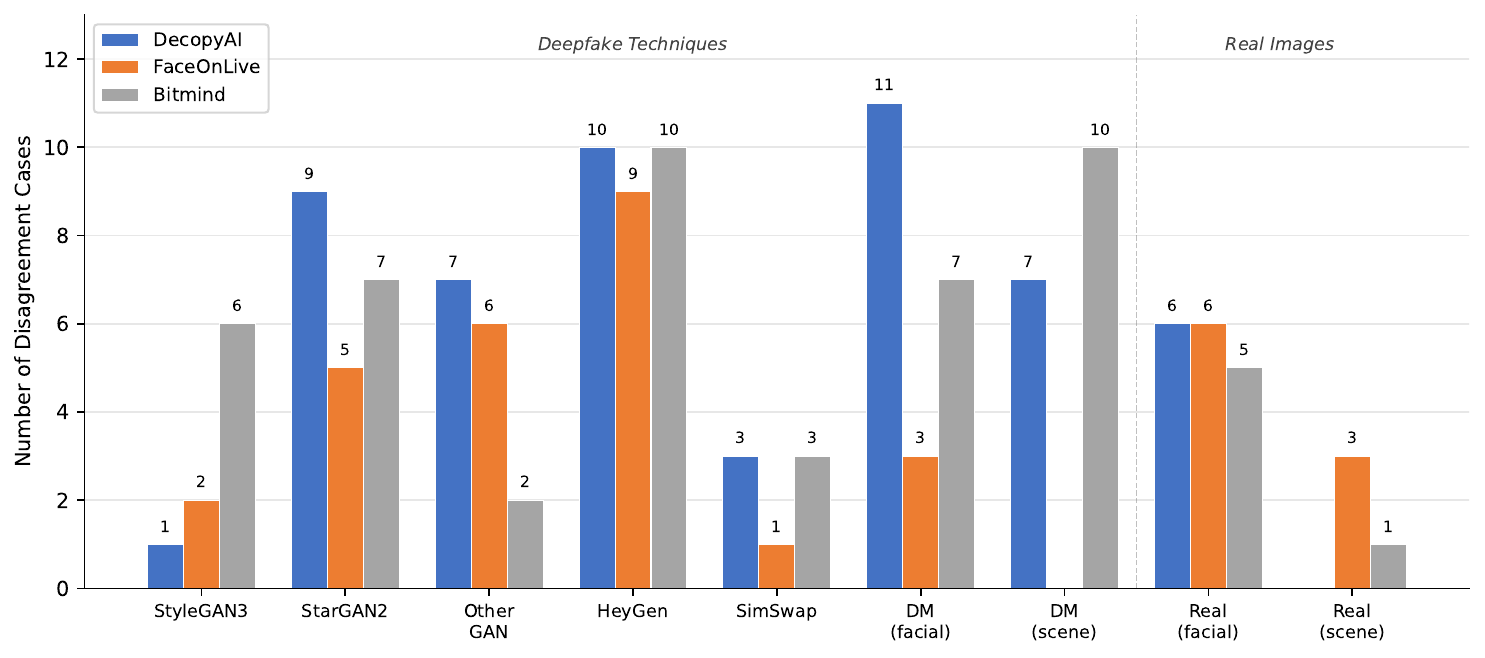}
    \caption{Number of human--AI disagreement cases across generative techniques for each AI classifier.}
\label{fig:discordance-technique}
\end{figure*}

Figure~\ref{fig:discordance-technique} shows the distribution of disagreement cases across generative methods. The most prominent pattern is the concentration of disagreements on HeyGen images, where all three tools classified every image as real. This aligns with the complete detection failure reported in Table~\ref{tab:generation} and confirms that HeyGen outputs fall entirely outside the classifiers' learned decision boundaries.

Among GAN-based methods, StarGAN2 produces the highest disagreement rates for DecopyAI and FaceOnLive, while Bitmind's disagreements distribute more evenly across techniques. For diffusion model outputs, FaceOnLive shows no disagreement on scene-category images (detecting all 20 correctly), whereas DecopyAI and Bitmind exhibit 7 and 10 disagreements respectively in this category. Disagreements on real images are rare across all tools, consistent with the high specificity (real detection rates $>$91\%) reported in the main paper.

These technique-level patterns reinforce the finding that classifier blind spots are generator-specific rather than uniform, and that the combination of multiple tools with human oversight is necessary to achieve reliable coverage across the full spectrum of contemporary generation methods.

\subsection{Confidence Scores Analysis}\label{appendix:confidence_analysis}

The aggregate accuracy metrics in Table~\ref{tab:overall_accuracy} reveal \textit{what} tools get wrong, but not \textit{how confidently} they get it wrong. This distinction matters for practical deployment: a tool that assigns low confidence to its errors provides an implicit warning for manual review, whereas a tool that expresses high confidence on incorrect predictions is actively misleading. This section examines whether the confidence scores produced by the three AI classifiers serve as reliable indicators of prediction correctness.

A prerequisite for comparison is aligning confidence semantics across tools. DecopyAI and FaceOnLive report the estimated probability of manipulation (higher = more likely fake), whereas Bitmind reports confidence in its binary verdict. For an image classified as real, Bitmind may report 95\% confidence, indicating strong certainty in the \textit{real} assessment rather than high DF probability. To enable uniform comparison, Bitmind's scores for real-classified images are inverted so that all values represent estimated DF probability.

\subsubsection{Aggregate Confidence by Classification Outcome}

Table~\ref{tab:confidence_outcomes} disaggregates average confidence scores across the four classification outcomes.

\begin{table}[h]
\centering
\caption{Average confidence scores (\%, as deepfake probability) for each classification outcome.}
\label{tab:confidence_outcomes}
\begin{tabular}{lccc}
\hline
 & DecopyAI & FaceOnLive & Bitmind \\
\hline
True Positives  & 88.29 & 94.79 & 86.00 \\
True Negatives  & 3.10  & 3.93  & 24.00 \\
False Positives & 94.02 & 75.60 & 79.93 \\
False Negatives & 3.91  & 7.26  & 24.27 \\
\hline
\end{tabular}
\end{table}

Correctly classified images behave as expected: high confidence on true positives and low confidence on true negatives. The critical finding concerns errors. DecopyAI assigns 94.02\% average confidence to false positives, which is \textit{higher} than its true positive average (88.29\%). This means that the tool is more confident when wrongly flagging real images than when correctly identifying DFs. FaceOnLive and Bitmind exhibit the same pattern at reduced magnitude. For false negatives, all three tools assign very low DF probability (3.91--24.27\%), expressing near-complete certainty that synthetic images are authentic. Practitioners receiving these outputs would have no basis to suspect tool failure.

\subsubsection{Confidence Score Distribution}

Table~\ref{tab:confidence_distribution} presents the confidence distribution across five bins, alongside the human baseline.

\begin{table}[h]
\centering
\caption{Distribution of confidence scores across five bins. Bins 1--5 correspond to 0--20\%, 20.01--40\%, 40.01--60\%, 60.01--80\%, and 80.01--100\% estimated deepfake probability, respectively.}
\label{tab:confidence_distribution}
\begin{tabular}{lccccc}
\hline
Bin & Range & Human & DecopyAI & FaceOnLive & Bitmind \\
\hline
1 & 0--20\%      & 29 & 97 & 75 & 5 \\
2 & 20.01--40\%  & 35 & 4  & 1  & 93 \\
3 & 40.01--60\%  & 15 & 4  & 7  & 0 \\
4 & 60.01--80\%  & 24 & 9  & 6  & 4 \\
5 & 80.01--100\% & 47 & 36 & 61 & 48 \\
\hline
\end{tabular}
\end{table}

DecopyAI and FaceOnLive exhibit strongly bimodal distributions, clustering at the scale extremes with minimal intermediate scores. Both tools function as near-binary classifiers that express either very high or very low confidence, providing no uncertainty signal. Bitmind's lower bin shifts toward 20--40\%, consistent with its higher baseline scores in Table~\ref{tab:confidence_outcomes}. The human evaluator, by contrast, distributes confidence across all five bins, reflecting calibrated assessment that acknowledges gradations of certainty where ambiguous cases concentrate.

\subsubsection{Confidence by Generative Technique}

Table~\ref{tab:confidence_technique} disaggregates confidence for correctly detected deepfakes (TP) and missed DFs (FN) by generative technique.

\begin{table}[h]
\centering
\caption{Average confidence scores (\%) for true positives (TP) and false negatives (FN) by generative technique. ``--'' indicates no cases for that tool-technique combination.}
\label{tab:confidence_technique}
\begin{tabular}{llcccc}
\hline
Technique & & DecopyAI & FaceOnLive & Bitmind \\
\hline
StyleGAN3    & TP & 92.52 & 91.00 & 86.40 \\
             & FN & --    & 23.00 & 24.44 \\
StarGAN2     & TP & 95.48 & 93.80 & 84.27 \\
             & FN & 0.28  & 3.00  & 22.84 \\
Other GAN    & TP & 54.65 & 91.67 & 86.63 \\
             & FN & 6.31  & 7.71  & 22.76 \\
HeyGen       & TP & --    & --    & -- \\
             & FN & 0.05  & 5.20  & 24.70 \\
SimSwap      & TP & 89.75 & 90.67 & 86.17 \\
             & FN & 8.61  & 48.00 & 29.10 \\
DM (facial)  & TP & 89.61 & 96.71 & 84.80 \\
             & FN & 8.06  & 1.33  & 25.81 \\
DM (scene)   & TP & 90.54 & 97.45 & 87.58 \\
             & FN & 3.18  & --    & 22.54 \\
\hline
\end{tabular}
\end{table}

The most consequential finding concerns HeyGen, the one generation method where all three classifiers failed entirely (Table~\ref{tab:generation}). Despite detecting zero HeyGen images, the tools express near-complete certainty in their incorrect assessments: DecopyAI averages 0.05\%, FaceOnLive 5.20\%, and Bitmind 24.70\% DF probability. A practitioner relying on these scores would receive unambiguous confirmation that every HeyGen image is authentic. A similar pattern appears for StarGAN2, where DecopyAI detects only 1 of 10 images but assigns 0.28\% average DF probability to the 9 misses. FaceOnLive shows slightly better calibration on SimSwap false negatives (48.00\%), approaching the decision boundary, but this remains isolated rather than systematic.

These results establish that current AI classifier confidence scores are poorly calibrated indicators of prediction reliability. The scores reinforce correct predictions but provide no warning when the tool is wrong. For investigative workflows, confidence scores cannot substitute for independent verification, and practitioners should treat automated predictions as one input among several rather than as calibrated probability estimates.

\subsection{Usability and Result Interpretability}

Detection accuracy alone does not determine a tool's practical value. A tool that delivers correct predictions but presents them ambiguously may be less useful than a slightly less accurate tool with clear, interpretable output. This section evaluates usability and feedback quality across both paradigms, combining a standardised quantitative assessment for the AI classifiers with qualitative observations for the forensic platforms.

\subsubsection{AI Classifier Usability}

The System Usability Scale (SUS)~\cite{lewis2018system} was applied to the three AI classifiers. SUS is a ten-item Likert-scale questionnaire yielding a score from 0 to 100, where scores above 68 are considered above average and scores above 80 indicate excellent usability. Table~\ref{tab:sus_responses} presents the full response set.

\begin{table}[h]
\centering
\caption{SUS responses (1\,=\,strongly disagree, 5\,=\,strongly agree). Odd items are positively worded; even items negatively worded. Dec.\,=\,DecopyAI; FoL\,=\,FaceOnLive; Bit.\,=\,Bitmind.}
\label{tab:sus_responses}
\footnotesize
\setlength{\tabcolsep}{4pt}
\begin{tabular}{@{}lccc@{}}
\hline
Statement & Dec. & FoL & Bit. \\
\hline
S1: Use frequently         & 3 & 4 & 4 \\
S2: Unnecessarily complex   & 1 & 2 & 1 \\
S3: Easy to use             & 5 & 4 & 5 \\
S4: Need technical support  & 1 & 1 & 1 \\
S5: Well integrated         & 4 & 3 & 5 \\
S6: Too much inconsistency  & 1 & 1 & 1 \\
S7: Learn quickly           & 5 & 5 & 5 \\
S8: Cumbersome              & 1 & 2 & 1 \\
S9: Felt confident          & 5 & 5 & 5 \\
S10: Needed to learn a lot  & 1 & 1 & 1 \\
\hline
\textbf{SUS Score}          & \textbf{92.5} & \textbf{85.0} & \textbf{97.5} \\
Clicks to result            & 2 & 3 & 3 \\
\hline
\end{tabular}
\end{table}

All three tools score well above the ``excellent'' threshold of 80. Bitmind achieves the highest score (97.5), reflecting its minimal interface. DecopyAI scores 92.5, with its two-click workflow contributing to perceived simplicity. FaceOnLive scores slightly lower (85.0) due to an additional analysis-initiation step and multiple sub-predictions (overall, deepfake-specific, face-manipulation-specific) that introduce minor cognitive overhead. All tools require only 2--3 clicks from image selection to result, making them accessible to non-technical users.

\subsubsection{Feedback Richness and Transparency}

High usability does not guarantee interpretable or actionable output. We evaluated each AI classifier against eight feedback quality criteria. Table~\ref{tab:feedback_richness} presents the results.

\begin{table}[h]
\centering
\caption{Feedback richness assessment. \checkmark\,=\,present; \texttimes\,=\,absent.}
\label{tab:feedback_richness}
\footnotesize
\setlength{\tabcolsep}{4pt}
\begin{tabular}{@{}lccc@{}}
\hline
Criterion & DecopyAI & FaceOnLive & Bitmind \\
\hline
Clear verdict             & \checkmark & \checkmark & \checkmark \\
Confidence score          & \checkmark & \checkmark & \checkmark \\
Textual explanation       & \texttimes & \texttimes & \texttimes \\
Annotations / heatmaps   & \texttimes & \texttimes & \texttimes \\
Easy to interpret         & \checkmark & \checkmark & \texttimes \\
Process described         & \texttimes & \texttimes & \checkmark \\
Help / documentation      & \texttimes & \checkmark & \texttimes \\
Next-step guidance        & \texttimes & \texttimes & \texttimes \\
\hline
\textbf{Total (/8)}      & \textbf{3} & \textbf{4} & \textbf{3} \\
\hline
\end{tabular}
\end{table}

No tool provides textual explanations, image annotations, heatmaps, or next-step recommendations. All deliver a verdict and confidence score, but none explains \textit{why} a prediction was made. Bitmind partially compensates by displaying sub-model confidence distributions, though its inverted confidence semantics (Appendix~\ref{appendix:confidence_analysis}) reduce interpretive clarity. FaceOnLive achieves the highest feedback score by providing accessible documentation and separating output into three prediction categories. DecopyAI offers a downloadable PDF report containing only the verdict and score. The consistent absence of explanatory feedback represents a significant limitation for investigative use: practitioners receive a binary label and a percentage but no evidence trail.

\subsubsection{Forensic Analysis Platform Usability}

Forensic platforms were not evaluated using SUS because their interactive, multi-algorithm workflows do not conform to the single-task interaction model SUS assumes. We report qualitative observations from the investigator's experience.

\textbf{FotoForensics} offers the simplest interface: upload an image and navigate pre-computed analysis tabs (ELA, hidden pixels, metadata). No parameters are adjustable and results appear within seconds, making it the fastest forensic tool for initial screening but also the most limited when results are ambiguous.

\textbf{Forensically} provides the greatest analytical depth through per-method sensitivity controls (sliding scales for minimal similarity, cluster size, block size, and other parameters). However, this configurability introduces a significant usability challenge: different sensitivity settings for the same image frequently produced contradictory results. The investigator must understand each detection method's mechanics to select appropriate parameters, making the learning curve steep and the time investment per image substantially higher.

\textbf{InVID \& WeVerify} offers the most extensive algorithm collection (ELA, WAVELET, DCT, Double Quantisation, CAGI, BLOCK, CMFD, Mantranet, FUSION) but no user-adjustable sensitivity. Its documentation acknowledges that individual detectors may flag suspicious activity in both authentic and manipulated images, recommending that confidence be drawn from convergence across multiple capabilities. In practice, this aggregation burden falls entirely on the investigator, contributing to the tool's 37.8\% real detection rate (Table~ \ref{tab:overall_accuracy}).

\subsubsection{Cross-Paradigm Comparison}

Table~\ref{tab:usability_comparison} summarises usability characteristics across paradigms.

\begin{table}[h]
\centering
\caption{Cross-paradigm usability comparison.}
\label{tab:usability_comparison}
\footnotesize
\setlength{\tabcolsep}{3pt}
\begin{tabular}{@{}lll@{}}
\hline
Dimension & Forensic & AI Classif. \\
\hline
Time/image        & Minutes          & Seconds \\
Expertise req.    & High             & None \\
Tunable params.   & Forensic. only   & None \\
Output format     & Visual overlays  & Binary + \% \\
Interp.\ burden   & On investigator  & Minimal \\
Transparency      & Full             & Opaque \\
Explan.\ feedback & Visual only      & None \\
Evid.\ trace      & High             & None \\
\hline
\end{tabular}
\end{table}

The two paradigms present an inverse usability--transparency trade-off. AI classifiers are highly usable (SUS 85--97.5) and require no expertise, but operate as opaque systems providing no interpretable evidence. Forensic platforms demand substantial expertise and time but offer full analytical transparency, enabling investigators to reason about \textit{why} an image may be manipulated. Neither paradigm alone satisfies the requirement for tools that are simultaneously accessible, transparent, and accurate, reinforcing the hybrid workflow recommendation in Section~\ref{sec:discussion-future}.

\end{document}